\documentstyle{article}
\def\Red{}
\def\Black{}
\def\Blue{}
\newcommand{\GeV}{\,{\rm GeV}}
\newcommand{\TeV}{\,{\rm TeV}}

\newcommand{\NPB}[3]{{\em Nucl. Phys. \bf B#1} (#2) #3}
\newcommand{\PRD}[3]{{\em Phys. Rev. \bf D#1} (#2) #3}
\newcommand{\PRL}[3]{{\em Phys. Rev. Lett. \bf #1} (#2) #3}
\newcommand{\PRA}[3]{{\em Phys. Rev. \bf A#1} (#2) #3}
\newcommand{\PL}[3]{{\em Phys. Lett. \bf #1} (#2) #3}
\newcommand{\Frac}[2]{\leavevmode\kern.1em\raise.5ex
\hbox{\the\scriptfont0 #1}
\kern-.1em/\kern-.15em\lower.25ex\hbox{\the\scriptfont0 #2}}
\makeatletter
\newcounter{alphaequation}[equation]
 \def\thealphaequation{\theequation\hbox to
0.6em{\hfil\alph{alphaequation}\hfil}}
\def\eqnsystem#1{
\def\@eqnnum{{\rm (\thealphaequation)}}
\def\@@eqncr{\let\@tempa\relax
\ifcase\@eqcnt \def\@tempa{& & &}
\or \def\@tempa{& &}\or \def\@tempa{&}\fi\@tempa
\if@eqnsw\@eqnnum\refstepcounter{alphaequation}\fi
\global\@eqnswtrue\global\@eqcnt=0\cr}
\refstepcounter{equation}
\let\@currentlabel\theequation
\def\@tempb{#1}
\ifx\@tempb\empty\else\label{#1}\fi
\refstepcounter{alphaequation}
\let\@currentlabel\thealphaequation
\global\@eqnswtrue\global\@eqcnt=0
\tabskip\@centering\let\\=\@eqncr
$$\halign to \displaywidth\bgroup
  \@eqnsel\hskip\@centering
  $\displaystyle\tabskip\z@{##}$&\global\@eqcnt\@ne
  \hskip2\arraycolsep\hfil${##}$\hfil&
  \global\@eqcnt\tw@\hskip2\arraycolsep
  $\displaystyle\tabskip\z@{##}$\hfil
  \tabskip\@centering&\llap{##}\tabskip\z@\cr}

\def\endeqnsystem{\@@eqncr\egroup$$\global\@ignoretrue}
\makeatother

\def\diag{\mathop{\rm diag}}
\newcommand{\mb}[1]{\mbox{\normalsize\boldmath $#1$}}

\def\SU{{\rm SU}}
\def\SO{{\rm SO}}

\def\circa#1{\,\raise.3ex\hbox{$#1$\kern-.75em\lower1ex\hbox{$\sim$}}\,}

\newcommand{\mysection}[1]{\setcounter{equation}{0}\section{#1}}
\renewcommand{\theequation}{\thesection.\arabic{equation}}

\begin{document}
July 1995\hfill\vbox{
\hbox{\bf IFUP -- TH-42-95}\hbox{\bf hep-ph/9507379}}\\[5mm]

\centerline{\huge\bf\Red Lepton Flavour Violations}
\centerline{\huge\bf in SO(10) with large $\tan\beta$}\bigskip\Black
\centerline{\large\bf
Paolo Ciafaloni\footnote{E-Mail: ciafalon@sunthpi1.difi.unipi.it},
Andrea Romanino\footnote{E-Mail: romanino@sunthpi1.difi.unipi.it}
and Alessandro Strumia\footnote{E-Mail: Strumia@ibmth.difi.unipi.it}}
\bigskip
\centerline{\large\em Dipartimento di Fisica, Universit\`a di Pisa}
\centerline{\large and}
\centerline{\large\em INFN, Sezione di Pisa, I-56126 Pisa, Italy}
\bigskip\bigskip\Blue
\centerline{\large\bf Abstract}
\begin{quote}\large\indent
In supersymmetric~SO(10) with large
$\tan\beta$ and Yukawa coupling unification
$\lambda_t=\lambda_b=\lambda_\tau$, we compute the rates for lepton
flavour violating processes. Experiments in progress or foreseen for
$\mu\to e\gamma$, $\mu\to e$ conversion (and the dipole moment of the
electron) are shown to provide very significant tests of the theory for
all slepton masses up to about $1\TeV$.
\end{quote}\Black

\mysection{Introduction}
The possibility to understand the rationale of the
quantum numbers of quarks and leptons strongly supports
the view of a local symmetry larger than the standard model one,
relevant at, or close to, the Planck scale.
In this context, a phenomenologically very relevant question arises:
is this larger symmetry already
present in the field theory which emerges below the Planck scale
and describes all the interactions but gravity?
Indeed, the measured values of the
standard model gauge couplings seem to favour the possibility of
an intermediate stage of gauge-unification,
around $M_{\rm G}\approx 2\cdot 10^{16}\GeV$.
The exploration of all possible signals of such unification
acquires therefore a great significance.
Among these signals, one which is especially
relevant in the case of supersymmetric unification,
is the violation of lepton flavour~\cite{BH}.

Soon after the first formulation of phenomenologically
viable supersymmetric models, it was realized that,
in absence of lepton flavour conservation, a
generic slepton mass matrix would give rise to uncontrollably large
lepton flavour violations (LFV)~\cite{EN}.
Without a theoretical guideline, it is however virtually impossible to
make a reliable prediction for the corresponding rates.
On one side the mixing angles in the leptonic sector are unknown; on
the other side, and even more importantly, there is in general no
control of the amount of non-degeneracy in the slepton masses, which is
essential to undo a GIM-like cancellation in the relevant amplitudes.
For lepton-slepton mixing angles
of the order of the Cabibbo-Kobayashi-Maskawa (CKM) ones,
the $\tilde{e}$ and $\tilde{\mu}$
sleptons must be extremely degenerate~\cite{EN,DS}
in order not to exceed the experimental upper bounds on LFV processes,
while a ${\cal O}(1)$ splitting between the
$\tilde{\tau}$ and the other sleptons may be acceptable.
To avoid a dangerous non degenerate sfermion spectrum,
a flavour universality hypothesis was made
on the soft SUSY breaking terms.
Moreover, this hypothesis has been made plausible in some models,
for example when SUSY breaking is
communicated to observable fields by gravity~\cite{BFS}.
In this case the universality holds at the Planck scale.

Remarkably enough, a controllable source for flavour non-universality
is present in minimal supersymmetric unified models, as pointed out by
Barbieri and Hall~\cite{BH}.
When the SUSY breaking parameters evolve to low energy,
their universality is lost if
non flavour-universal couplings are present.
A significant splitting between the $\tilde{\tau}$ mass
and the $\tilde{e}$ and $\tilde{\mu}$ ones
is induced by the top Yukawa coupling
(and, possibly, the other third generation couplings).
As a consequence, the rates for LFV processes can be predicted
in the usual parameter space of the MSSM.
As found in~\cite{8Art}, these rates
are of key interest as tests of supersymmetric unification.

The hierarchy between the third generation fermion
masses can be interpreted as a hierarchy between their Yukawa couplings.
In this case $\lambda_{\tau}$
cannot influence significantly
the sleptons masses and the resulting LFV rates were
computed in~\cite{8Art}.
It is the purpose of the present work to extend
this analysis to theories where the $t/b,\tau$ splitting is instead
attributed to a hierarchy
between the vacuum expectation values of the two Higgs doublets,
or, in the usual language, to a large
$\tan\beta\equiv v_{\rm u} / v_{\rm d} \sim m_t/m_b$.
This possibility is particularly interesting since it allows for
a~SO(10) GUT where not only the gauge couplings but also the third
generation Yukawa couplings unify at $M_{\rm G}$~\cite{ALS,tbu}.
In this paper we compute, in the large $\tan\beta$ region,
the minimal amount of flavour violations
that in a wide class of supersymmetric unified models
we expect to be transferred
from the hadronic to the leptonic sector
by the unified third generation Yukawa coupling.

In section~\ref{Model} we present and motivate the model in which we
calculate the lepton flavour violations.
We explain in which sense this model gives the minimal amount
of lepton flavour violating effects we expect to be present
in a general class of unified models.
In section~\ref{Spettro} we show how the well known problems of
the large $\tan\beta$ region are solved by an unification scale
$D$-term contribution to the soft SUSY breaking masses,
which also fixes the main features of the spectrum.
In section~\ref{LFV} we show our predictions for the
lepton flavour violating processes, and
in section~\ref{TeXtures} we discuss how they
are expected to be modified in more general models.
In appendices~A and~B we give and solve
the renormalization group equations
(RGEs) above and below the unification scale.

\mysection{The model}\label{Model}
We restrict our attention to unified theories with SO(10) gauge group,
because they furnish the main (unique?) motivation for considering a
large $\tan\beta\sim m_t/m_b$ value as an interesting one,
notwithstanding the difficulties in obtaining it in a radiative
electroweak symmetry breaking scenario.

To exploit the full predictability of SO(10) gauge unified theories
(GUT), we only consider softly broken supersymmetric
field theory models, assumed to be valid up to the
Planck scale, in which
\begin{enumerate}
\item the two MSSM Higgs doublets, $h^{\rm u}$ which gives mass to
up-quarks and $h^{\rm d}$ which gives mass to down-quarks
and to leptons, lie in a single 10 (`$\Phi$') vector representation of
SO(10);
\item the third generation particles lie in a single $16$ (`$\Psi_3$')
spinorial representation of SO(10), together with a right handed
neutrino;
\item the only relevant couplings for lepton flavour violations are the
unified gauge coupling and the unified third generation Yukawa coupling;
\item the light generation Yukawa couplings are supposed to arise from
non renormalizable operators, allowing in this way for non trivial
physical flavour-mixing angles
$$f=\lambda \Psi_3\Psi_3\Phi + \hbox{n.r.o};$$
\item the soft SUSY-breaking scalar masses are supposed to be
{\em universal at the Planck scale\/}\footnote{This assumption is done
in order to reduce the number of free parameters
and is perfectly consistent with our intention of
compute the minimal amount of LFV.
In fact, apart from accidental cancellations, possible flavour violations
already present in the soft masses at the Planck scale would simply
add to the renormalization induced ones, while non universal but
flavour-symmetric soft breaking masses would not substantially alter
the LFV rates.}.
In view of the different origin of the various Yukawa couplings, it
would not be satisfactory to make a similar hypothesis for the trilinear
$A$-terms --- and we will not do it;
\item the unification gauge group SO(10) directly breaks to the
standard model gauge group at the unification scale.
The $D$-term contribution to the soft SUSY breaking scalar masses
corresponding to the reduction of the rank of the gauge group should
be non zero\footnote{This can be achieved without introducing new
sources of LFV.} in order to minimize the fine-tuning
necessary for a correct electroweak symmetry breaking.
The unified gauge $\beta$-function coefficient is not an
important parameter~\cite{8Art}.
\end{enumerate}
Before being able to compute the LFV effects in this
model we should solve two new kind of problems --- one of technical and
one of fundamental nature --- both peculiar of the large $\tan\beta$
region.
To be able to calculate the rates for the LFV processes, one has
to control both the
$m_{\tilde{\tau}}/m_{\tilde{\mu}},m_{\tilde{e}}$ splitting
and the lepton-slepton mixing angles.
In previous calculations of LFV effects for
moderate values of $\tan\beta$,
two different and independent sources
produced these two effects:
the mass splitting was induced by
the (diagonalizable) up quark Yukawa couplings matrix above the
unification scale, while the mixing angles resulted from the ones in
the lepton Yukawa matrix, linked by unification physics to the
CKM angles.

The technical problem is that now, below the unification scale,
also the $\tan\beta$ enhanced $\tau$ Yukawa coupling contributes to the
intergenerational slepton mass splitting --- this is discussed in
appendix~B.
This coupling also gives some reduction of the lepton-slepton mixing
angles, both in the left and in the right sector.
Appendix~\ref{VeEvol} is devoted to this computation.

The fundamental problem is that now, above the unification
scale, all the Yukawa couplings necessary to give mass to the light
fermions, may also induce lepton-slepton mixing angles.
We cannot say anything about them, without understanding
the dynamical mechanism that generates the light generation Yukawa
couplings. However, if we
insist on computing only the 'minimal'  effects, we may (and we
shall) assume that the main consequence of renormalization effects
above the unification scale consists in making the third generation
sfermions in the $\Psi_3$ lighter than the other ones, {\em
without\/} generating non-diagonal entries in the slepton mass matrices.

In this case the lepton-slepton mixing angles only come from
the lepton Yukawa coupling matrix, and
we have to connect them with the CKM angles.
These angles, which measure the misalignment between the up and the down
quark angles in the left sector, are the only experimentally accessible
angles below the Fermi scale.
We will again assume, as in the previous analyses, that the down-quark
and lepton Yukawa matrices are equal at the unification scale.
However the mixing angles in the down-quark Yukawa matrix  may now only
be a partial contribution to the measured CKM ones, with the other
contribution coming from a non diagonal up-quark Yukawa matrix.
The unknown up-quark mixing angles, if comparable in size to the CKM
ones, could generate an electric dipole for the $u$ quark giving in this
way a contribution of the same order of the $d$ quark one to the neutron
electric dipole moment.
Due to the larger mass hierarchy among the up-quarks, it looks however
more likely that the mixing angles in the up sector be smaller than the
ones in the down sector.
We stick to this simplifying hypothesis in the following, postponing to
section~\ref{TeXtures} a discussion of how we may expect the mixing
angles to be distributed between the up, the down and lepton Yukawa
coupling matrices.

\begin{figure}[t]\setlength{\unitlength}{1cm}
\begin{center}
\begin{picture}(16,6)
%\put(-0.5,0){\special{picture alpha3}}
%\put(8.5,0){\special{picture 16/10}}
\put(-0.5,0){\includegraphics{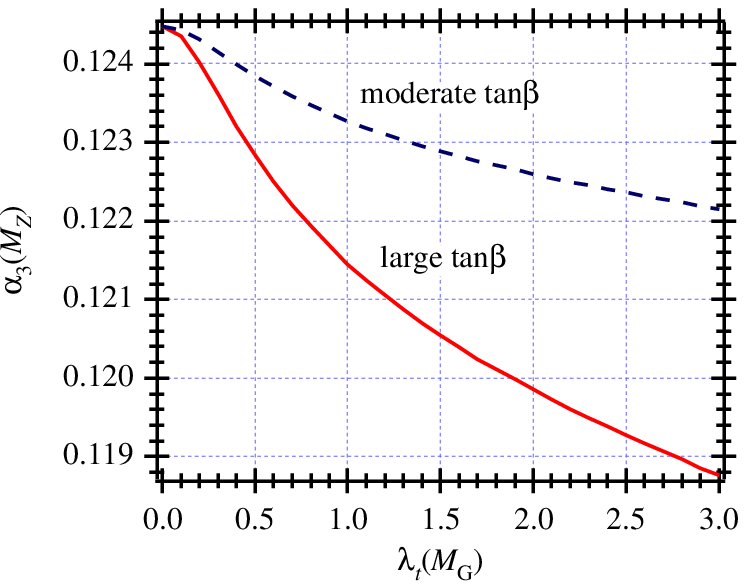}}
\put(8.5,0){\includegraphics{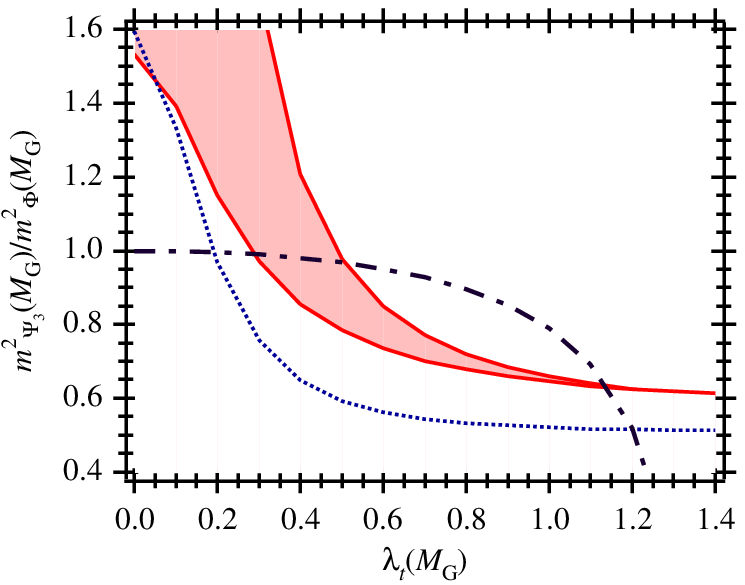}}
\end{picture}
\parbox[b]{8cm}{\caption{Two loop prediction for $\alpha_3(M_Z)$ as
function of the top quark
Yukawa coupling at the unification scale,
$\lambda_{t\rm G}$, in the cases of moderate $\tan \beta$,
$\lambda_{t\rm G} \gg \lambda_{b\rm G},\lambda_{\tau\rm G}$ (dashed
line), and large $\tan\beta$,
$\lambda_{t\rm G} = \lambda_{b\rm G}=\lambda_{\tau\rm G}$ (solid line),
with all threshold and gravitational effects set to
zero.\label{fig:alpha3}} }
\hfill
\parbox[b]{8cm}{\caption{The allowed interval for
$m_{\Psi_3\rm G}^2/m_{\Phi\rm G}^2$ (gray area) and
the prediction for its value from Planck scale
universality (dot-dashed line)
as a function of $\lambda_{\rm G}$ in the low
fine-tuning region of the soft SUSY breaking parameters.
Also shown is the sub-dominant
limit from the $\tilde{L}_3$ mass
(dotted line).
\label{fig:16/10}} }
\end{center}\end{figure}

We can now summarize the additional hypotheses that complete the
description of the model. At the unification scale, in the
supersymmetric basis (in which there are no flavour violations at the
gaugino vertices)
\begin{itemize}
\item[7.] the sfermion mass matrices are flavour-diagonal, that is
\begin{equation}
\mb{m}_{\Psi}^2 = \diag(m_{\Psi_1}^2, m_{\Psi_1}^2, m_{\Psi_3}^2).
\end{equation} The expressions for these masses in term of the universal
soft breaking parameters at the Planck scale are given in Appendix~A;
\item[8.] the down and lepton Yukawa coupling matrices
$\mb{\lambda}^{\rm d}$ and $\mb{\lambda}^{\rm e}$, {\em equal \rm and
\em symmetric\/}, are the only non flavour-diagonal matrices present in
the theory.
\end{itemize}
With these assumptions we are now ready to start the actual
calculations.
We will however first recall the constraints on $\lambda_{\rm G}$,
the common value of the third generation
Yukawa couplings $\lambda_t$,
$\lambda_b$ and $\lambda_\tau$ at the unification scale.
This parameter plays a crucial role in the determination of
the LFV rates.
We fix the $\tan\beta$ value as a function of
$\lambda_{\rm G}$ by requiring the $\tau$ mass $m_\tau$ to have its
measured value
$$\tan\beta\simeq\frac{v\lambda_\tau(M_Z)}{m_\tau/\eta_\tau} =
97\lambda_\tau(M_Z)$$
where $v=174\GeV$ and $\eta_\tau=0.986$ is the QED renormalization for
the $\tau$ mass between $m_\tau$ and the $Z$-pole.
It is interesting to compute the reduction in the $\alpha_3(M_Z)$
value in the $\overline{\rm MS}$ scheme, as predicted from gauge
coupling unification, due to the Yukawa terms in the two loop RGEs
between the Fermi and the unification scale.
This is shown in fig.~\ref{fig:alpha3}, where the $\alpha_3(M_Z)$
reference value in the case of zero Yukawa couplings is the one obtained
with all the unknown threshold
and gravitational corrections~\cite{Hall} set to
zero and with the input values of the electroweak couplings used
in~\cite{LP}.
We see that, for ${\cal O}(1)$ values of the
top Yukawa coupling at the unification scale,
in the large $\tan\beta$ region this effect gives a reduction in the
predicted $\alpha_3(M_Z)$ of order of its present $1\sigma$
experimental error;
this reduction becomes three times smaller
for moderate values of $\tan\beta$.

The third generation unified Yukawa coupling $\lambda_{\rm G}$ is
a very important parameter for lepton flavour violations too.
There exist two almost equivalent upper bounds on it.
The maximum value that $\lambda_{\rm G}$ can assume without developing a
Landau pole below the Planck scale at
$2.4\cdot10^{18}\GeV$ is around $\lambda_{\rm G}\circa{<} 1.4$.
The second upper bound arise from the correct electroweak symmetry
breaking requirements, that we will discuss in the next section.
There are also two lower bounds on $\lambda_{\rm G}$.
To obtain a top quark mass in the CDF range~\cite{CDF},
$\lambda_{\rm G}\circa{>} 0.5$ is needed~\cite{ALS,tbu}.
In this case all the third generation
Yukawa couplings at the $Z$-pole are greater than about
$\Frac{3}{4}$ of their maximum possible values (``{\em InfraRed fixed
point\/}''), which are, for $\alpha_3(M_Z) = 0.121$,
\begin{equation}
\lambda_t^{\rm max}(M_Z) = 1.06,\qquad
\lambda_b^{\rm max}(M_Z) = 0.99,\qquad
\lambda_\tau^{\rm max}(M_Z) = 0.64.\qquad
\end{equation} This gives a $\tan\beta$ value in the range $45\div 60$.

To predict the correct value of the $b/\tau$ mass ratio even higher
values of $\lambda_{\rm G}$ seem indeed to be necessary.
There may be a conflict between this requirement and the upper bounds
previously mentioned.
One should not forget, however, the various uncertainties
that can affect the $b/\tau$ mass prediction:
\begin{itemize}
\item sizeable dependence on the value of $\alpha_3(M_Z)$ in the range
$0.110\div 0.125$;
\item $\tan\beta$ enhanced one loop quantum correction to the bottom
mass~\cite{tbu};
\item second-third generation mixing contribution to the $b$ and/or
$\tau$ masses~\cite{Ross};
\item mixing between third generation particles and other heavy ones,
induced by SU(5)-breaking vacuum expectation values~\cite{DP};
\item a possible right-handed tau neutrino Yukawa coupling effect in the
RGEs, if its mass is lower than the unification mass as cosmological
and phenomenological considerations may indicate~\cite{Ross,Rattazzi}.
\end{itemize}
For these reasons, in the following, we will avoid to impose a correct
$b/\tau$ mass ratio, and neglect all the problems related to it.

\mysection{Structure of the allowed parameter space}\label{Spettro}
In this section we will discuss for which values of the SUSY breaking
parameters it is possible to obtain a phenomenologically acceptable
theory.
The reader may want to skip this section and jump
directly to the following one, where we present the predictions on the
$\mu\to e\gamma$ branching ratio,
that constitute the principal aim of this paper.

It is well known that the price we have to pay in order to have Yukawa
unification, is the need to fine tune the parameters which
determine $\tan\beta$ and $M_Z$ through the minimization conditions
of the MSSM potential
\begin{equation}\label{eq:Min}
\frac{2\mu  B}{\mu_{\rm u}^2 + \mu_{\rm d}^2} =
\sin2\beta\qquad{\rm and}
\qquad
\frac{\mu_{\rm u}^2\tan^2\beta-1\mu_{\rm d}^2}{\tan^2\beta-1}= -{M_Z^2
\over2}.
\end{equation}
In the large $\tan\beta$ region these equations become
\begin{equation}\label{eq:MinBigTan}
\frac{\mu  B}{\mu_{\rm u}^2 + \mu_{\rm d}^2}\approx
\frac{1}{\tan\beta}\qquad{\rm and}\qquad
\mu_{\rm u}^2 \approx -{M_Z^2 \over2}.
\end{equation}
The first minimization condition
requires $\mu$ and/or $B$ to be much smaller than the
pseudoscalar Higgs mass
$m_A^2 = \mu_{\rm u}^2 + \mu_{\rm d}^2$.
Since $B$ gets sizeable renormalization corrections from the
trilinear $A$-terms and
from the gaugino masses (see appendix~B), and $m_A^2$ is
generated from the SUSY breaking scalar masses,
the best way to satisfy
the large $\tan\beta $ condition in~(\ref{eq:MinBigTan}) with
the minimal amount of fine tuning
is to restrict our analysis to the case where the scalar masses
are the dominant soft SUSY-breaking parameters~\cite{tbu,Rattazzi}.

The problem is that, since the RGEs for $\mu_{\rm u}^2$ and
$\mu_{\rm d}^2$ are similar and the two Higgs doublets come from
the same~SO(10) multiplet, it is
difficult to reconcile the two requirements that
\begin{itemize}
\item $\mu_{{\rm u}}^2$ must be negative to break the electro-weak gauge
symmetry as in~(\ref{eq:MinBigTan}), while
\item $\mu_{{\rm d}}^2$ must be positive enough so that
the squared pseudoscalar Higgs mass is positive.
\end{itemize}
This conciliation is impossible if we assume that the only relevant
soft SUSY breaking term is an universal scalar mass $m_0^2$,
either at the unification scale~\cite{tbu},
or at the Planck scale.
In these cases the renormalization effects may indeed induce a
small difference between the two  Higgs squared masses, but only in
the bad direction
$\mu_{{\rm u}}^2 \ge \mu_{{\rm d}}^2$.
To reconcile the two requirements it is possible to move to the
`hard fine tuning' region where the GUT scale gaugino mass is
larger than the scalar masses.
Other possible solutions consist in relying on positive
{\em ad hoc\/} large
${\cal O}(20\%-30\%)$ GUT threshold corrections to the ratios
of SO(10)-linked quantities, like
$\lambda_t^2/\lambda_b^2$, or $m_{\tilde{u}_R}^2/m_{\tilde{d}_R}^2$, or
directly to
$\mu_{{\rm d}}^2/\mu_{{\rm u}}^2$~\cite{OP}.
All these corrections would however introduce
new uncontrolled uncertainties.

The best solution is given by the well known fact that, at the scale
where the rank of the gauge group is reduced by spontaneous breaking,
possible additional contributions to the soft SUSY breaking scalar
masses arise from the $D$-terms associated with the broken diagonal
(Cartan) generators. These contributions are generated whenever the
soft SUSY-breaking masses of
the fields whose vacuum expectation values reduce the rank of the
group are different.

In the $\SO(10)\to \SU(3)_c\otimes\SU(2)_L\otimes{\rm U}(1)_Y$ case,
even with all the SUSY breaking scalar
masses degenerate at the Planck scale,
an amount of non-degeneracy may be produced by
different interactions of the
$16_H$ and $\overline{16}_H$ fields
whose vacuum expectation values reduce the rank of the group.
In this way a non vanishing SU(5)-invariant correction to the soft SUSY
breaking scalar masses arises from the
$D$-term associated with the broken ${\rm U}(1)_X$ generator.
Decomposing the SO(10) fields in SU(5) multiplets
$$\Phi = \bar{5}\oplus 5,\qquad
\Psi_i = 10_i\oplus \bar{5}_i \oplus 1_i$$
these corrections modify the matching conditions at the
unification scale in the following, very precise, way:
\begin{equation}\label{eq:Dterm}
\begin{array}{ll}
   m_{\bar{5}}^2 = m_\Phi^2+2 m_X^2
(= m_{h^{\rm d}}^2)&
   m_5^2 = m_\Phi^2-2 m_X^2
(= m_{h^{\rm u}}^2 )\\[2mm]
   m_{\bar{5}_i}^2  = m_{\Psi_i}^2-3m_X^2
(= m_{\tilde{L}_i}^2 = m_{\tilde{d}_{Ri}}^2)&
   m_{10_i}^2 =m_{\Psi_i}^2+m_X^2
(=  m_{\tilde{Q}_i}^2 = m_{\tilde{e}_{Ri}}^2= m_{\tilde{u}_{Ri}}^2)
\end{array}
\end{equation}
where $m_X^2$ is the only additional parameter.
These corrections enter the RGE equations from the
unification scale to the Fermi one in a very peculiar way
(see appendix~B):
\begin{itemize}
\item these new $m_X^2$ contributions and the old usual SUSY breaking
parameters evolve in an independent way;
\item in a very good approximation the ${\rm U}(1)_X$ $D$-term
corrections to the masses are not renormalized.
\end{itemize}
We now explain how all the main features of the low energy spectrum
emerge in a very clear way from the analytical approximation
presented in appendix~\ref{Ianal}.
The full numerical calculation is only necessary to
get the details.

The low energy Higgs mass parameters may be written as
\begin{eqnsystem}{sys:udmass}
\mu_{\rm u}^2 \equiv m_{h^{\rm u}}^2 + \mu^2 & \approx &
m_{\Phi\rm G}^2 +\mu^2 + x_2^h M_{5\rm G}^2 -
9I_t - 2 m_X^2\\
\mu_{\rm d}^2 \equiv m_{h^{\rm d}}^2 + \mu^2 & \approx &
m_{\Phi\rm G}^2 +\mu^2 + x_2^h M_{5\rm G}^2 -
9I_b + 2 m_X^2
\label{and1}
\end{eqnsystem}
where $x_2^h\approx 0.53$ and
$I_t\approx I_b$
is the Yukawa coupling induced
renormalization group correction,
approximately equal for the two Higgs doublets
(see appendix~\ref{Ianal} for an analytical expression).
We may neglect this small difference, because we
rely on a positive $m_X^2\approx (m_A^2 + M_Z^2)/4$
to obtain the desired
splitting between the two Higgs mass parameters.

Is there any limitation to this way of obtaining the desired symmetry
breaking pattern?
We see from eq.~(\ref{eq:Dterm}) that a positive $m_X^2$ term also has
the dangerous effect of decreasing the squared masses
of the sparticles in the three $\bar{5}_i$ of SU(5).
Above some value of $m_X^2$, one of them will become negative.
The dominant upper bound on $m_X^2$ will be given by the
third generation masses,
because they are further reduced by Yukawa effects in the
renormalization.
The $m_{\tilde{b}_R}^2>0$ constraint will dominate at low values of the
gaugino masses, because
the squarks get larger Yukawa couplings than the sleptons.
Above some value of $M_2$ (around $m_{\tilde{e}_R}/5$)
the dominant constraint will become $m_{\tilde{L}_3}^2>0$.
This is because the squarks get positive corrections to their masses
from the gluinos, corrections that are bigger than those that the other
gauginos give to the slepton masses.

\begin{figure}[t]
\setlength{\unitlength}{1cm}
\begin{center}
\begin{picture}(17,8.3)
\put(-2,-1.7){\includegraphics{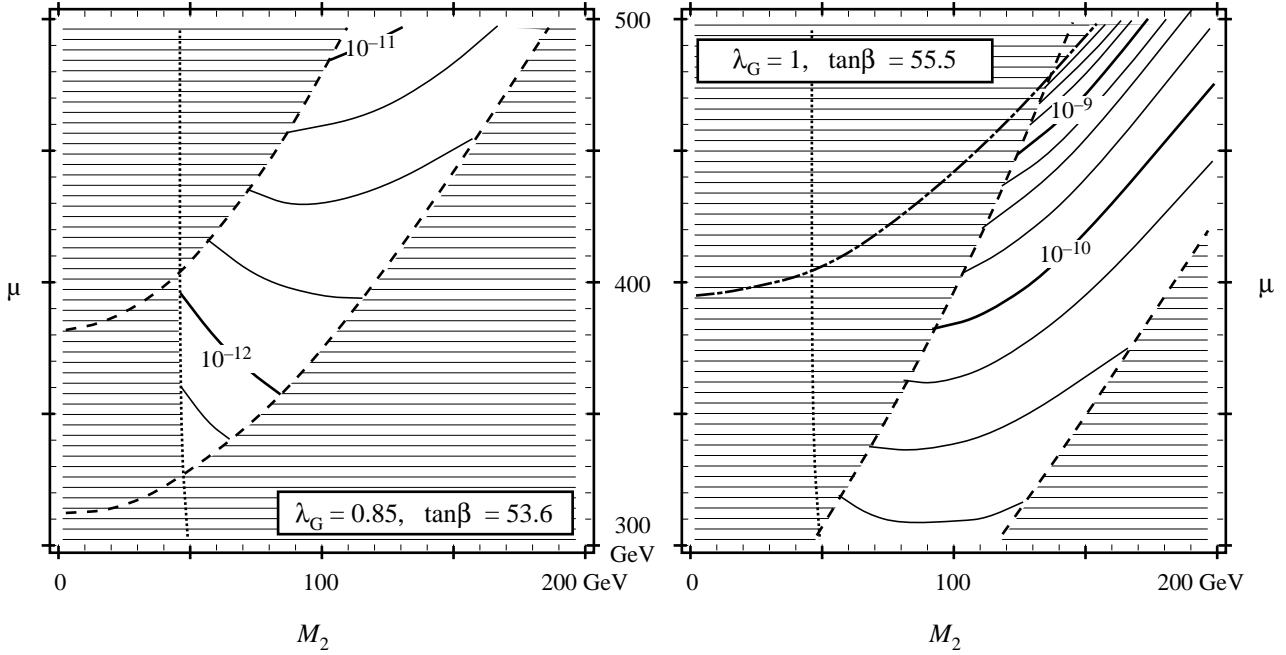}}
%\put(-0.2,0){\special{picture LFV}}
\end{picture}
\end{center}
\caption{Contour-plots for ${\rm B.R.}(\mu\to e\gamma)$
in the plane $\{M_2,\mu\}$ for
$m_{\tilde{e}_R}=1\TeV$ and
$\lambda_{\rm G} = \{0.85,1\}$.\label{fig:LFV}}
\end{figure}

Let us first understand these bounds in the limit where the soft scalar
masses, parametrized by
$m_{\Phi\rm G}^2$ and
$m_{\Psi_i\rm G}^2$, much larger than $M_Z^2$,
are the only non zero SUSY breaking parameters.
The dominant constraints $m_A^2>0$
and $m_{\tilde{b}_R}^2>0$
restrict the ratio $m_{\Psi_3\rm G}^2/m_{\Phi\rm G}^2$
in a $\lambda_{\rm G}$-dependent way.
 From our analytical approximation we get
$$\frac{21 - 5\rho_t}{2( 7 + 5\rho_t)}
\circa{<}\frac{m_{\Psi_3\rm G}^2}{m_{\Phi\rm G}^2}
\circa{<}\frac{7}{6\rho_t}-\frac{1}{2},\qquad{\rm where}\quad
\rho_t\approx\frac{\lambda_t^2(M_Z)}{(1.05)^2}.$$
This agrees very well with the full numerical calculation,
shown in fig.~\ref{fig:16/10},
from which we can see that, in this limit,
the allowed area closes at $\lambda_{\rm G}\simeq 1.2$.
In fig.~\ref{fig:16/10} we have also shown the predicted value for
the mass ratio assuming universality at $M_{\rm Pl}$
$$\frac{m_{\Psi_3\rm G}^2}{m_{\Phi\rm G}^2} =
\frac{1-\frac{15}{14}\rho_{\rm G}}{1-\frac{12}{14}\rho_{\rm G}},
\qquad{\rm where}\quad
\rho_{\rm G} =\frac{\lambda_{\rm G}^2}
{\lambda^2_{\rm max}(M_{\rm Pl})}.$$
It is now easy to understand what happens when we turn on small non zero values
for
the other SUSY-breaking parameters.
This is the case we are most interested in.
We fix the universal scalar mass
at some high value, for example imposing
$m_{\tilde{e}_R}^2 = 1\TeV$ as in the LFV plots of the next section,
and we examine the spectrum as a function of
the remaining parameters.
Its main characteristic are the following.
\begin{quote}
For a fixed ${\cal O}(1)$ value of $\lambda_{\rm G}$
and any value of the (almost irrelevant) $A$-parameters
the allowed region in the $\{M_2,\mu\}$ plane is the strip
\begin{equation}\label{eq:lowBound}
\mu^2 \approx(5.6\rho_t-2.2\rho_t^2-0.5) M_{5\rm G}^2
+ \Delta^2
\end{equation}
which becomes narrower for higher values of $\lambda_{\rm G}$.
Along its `lower' boundary, given by the $m_A^2>0$ condition,
the pseudoscalar and charged Higgs are light.
Along its `upper' boundary, given
by the $m_{\tilde{L}_3}^2>0$ ($m_{\tilde{b}_R}^2>0$) constraint
for values of $M_2$ greater (smaller) than about $m_{\tilde{e}_R}/5$,
a light $\tilde{\tau}_L$ ($\tilde{b}_R$) is present.
In the intermediate allowed region the
masses of these particles do not exceed
at the same time $\Frac{1}{3}$ of $m_{\tilde{e}_R}$.
\end{quote}
This behaviour may be seen from the LFV plots in fig.~\ref{fig:LFV},
in which the full numerical solutions have been employed.
Its features  can also be understood from the analytic approximation.
The dashed area corresponds to the excluded region.
The lower $m_A^2>0$ limit of~(\ref{eq:lowBound})
is easily obtained from eq.~(\ref{sys:udmass}) using the
analytic approximation for $I_t$, eq.~(\ref{sys:Iapprox}).
The fact that
the upper $m_{\tilde{L}_3}^2>0$ border is almost parallel to it
is not an accident and, again, is easily understood from the
analytic approximation.
The reason is that the dependence on the gaugino mass
and on the $\mu$ term of the doublet slepton mass
is dominated by the $-3m_X^2$ term,
which contain a ${\cal O}(M_3^2)$ term from $I_t$.
The other renormalization effects are negligible in comparison to it.

The origin of the $\Delta$ term in eq.~(\ref{eq:lowBound}) is also
easily understood from fig.~\ref{fig:16/10}.
For values of $\lambda_{\rm G}$ that give the correct value for the
$m_{\Psi_3\rm G}^2/m_{\Phi\rm G}^2$ ratio,
the strip starts from zero values of $M_2$ and $\mu$.
In the most interesting
region, $\lambda_{\rm G}\in 0.6\div1.1$,
the predicted ratio is somewhat high,
and the strip starts from a non-zero value of $|\mu|$
(typically $|\mu|>\Delta\sim m_{\tilde{e}_R}/5$);
in the remaining case $\lambda_{\rm G}\circa{>}1.1$
a non-zero value of $M_2$ will be instead necessary.

\begin{figure}[t]
\setlength{\unitlength}{1cm}
\begin{center}
\begin{picture}(17,10)
\put(-0.5,0){\includegraphics{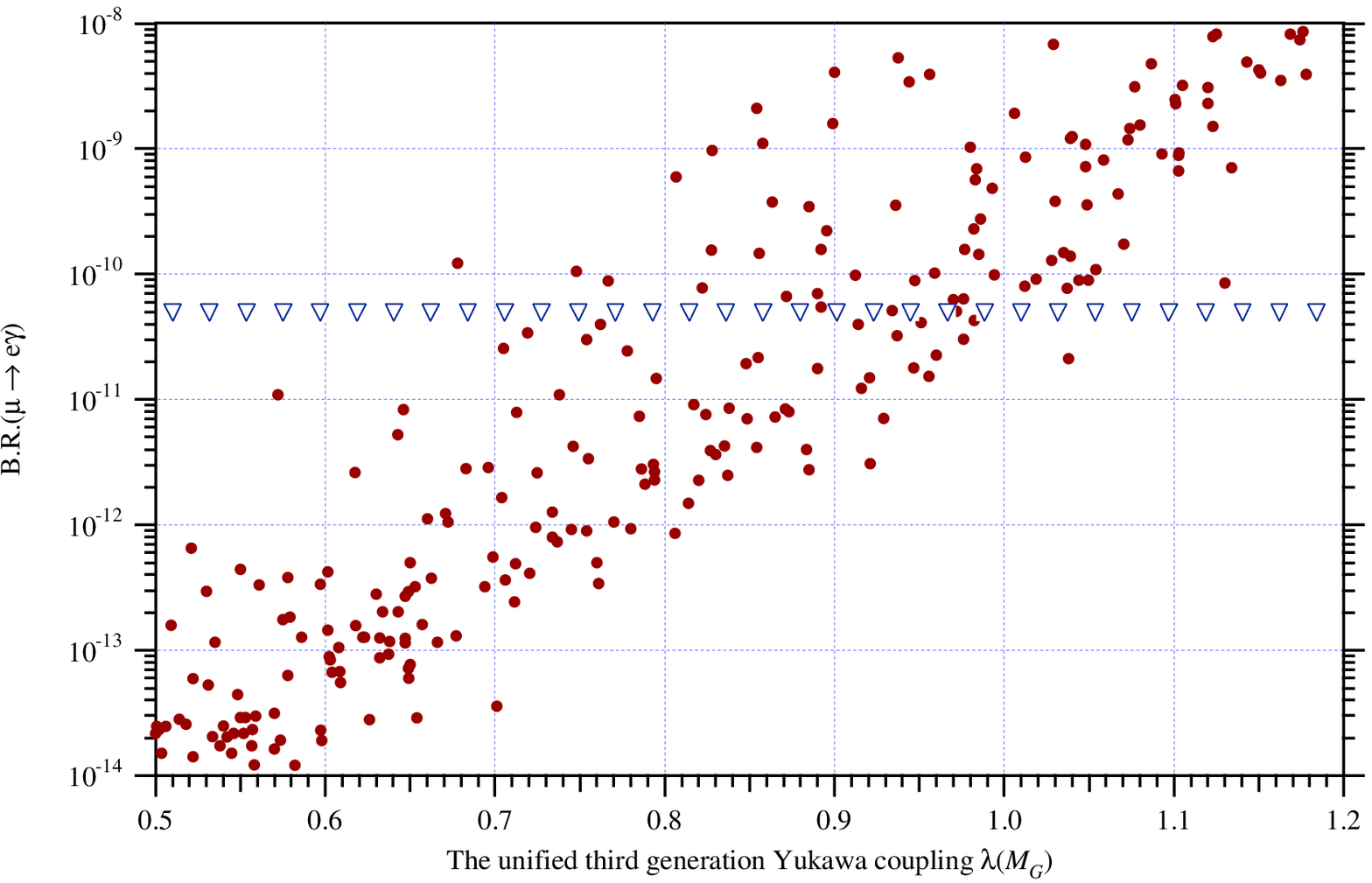}}
%\put(0,0){\special{picture BRlG}}
\end{picture}
\end{center}
\caption{${\rm B.R.}(\mu\to e\gamma)$
as function of $\lambda_{\rm G}$ for
$m_{\tilde{e}_R}=1\TeV$ and acceptable
values of the other free parameters.
The line of `$\bigtriangledown$' denotes
the experimental upper bound.\label{fig:BRlG}}
\end{figure}

\mysection{Leptonic flavour violations}\label{LFV}
Let us briefly recall
from~\cite{8Art} the main structure of the lepton flavour violations.

The most significant observables are the $\mu\to e\gamma$ decay, the
$\mu\to e$ conversion in atoms and the electric dipole moment $d_e$
(included in the list together with the genuine lepton flavour
violations because generated by the same mechanism~\cite{HD}).

In SO(10)-like theories, where LFV are mediated by both the left and
the right handed sleptons, a unique loop integral $F$,
with dimensions mass$^{-2}$, gives the
dominant contribution to all these processes
\begin{eqnsystem}{sys:LFV} {\rm B.R.}(\mu\to e\gamma) &=&
5.0\cdot 10^{-12}\times
\frac{|V^{{\rm e}_R}_{e\tilde{\tau}}V^{{\rm e}_L}_{\mu\tilde{\tau}}|^2+
|V^{{\rm e}_R}_{\mu
\tilde{\tau}}V^{{\rm e}_L}_{e\tilde{\tau}}|^2}
{2\cdot(0.01\cdot 0.04)^2}\frac{|F|^2}{(1\TeV)^{-4}}\\
{\rm C.R.}(\mu\to e~{\rm in~Ti}) &=&2.5\cdot 10^{-14}\times
\frac{|V^{{\rm e}_R}_{e\tilde{\tau}}V^{{\rm e}_L}_{\mu\tilde{\tau}}|^2+
|V^{{\rm e}_R}_{\mu\tilde{\tau}}V^{{\rm e}_L}_{e\tilde{\tau}}|^2}
{2\cdot(0.01\cdot 0.04)^2}\frac{|F|^2}{(1\TeV)^{-4}}\\
d_e &=& 2.9\cdot10^{-27}\,e\cdot {\rm cm}\times
\frac{|V^{{\rm e}_R}_{e\tilde{\tau}}V^{{\rm e}_L}_{e\tilde{\tau}}|}
{(0.01)^2}\frac{|F|}{(1\TeV)^{-2}}\sin\varphi
\end{eqnsystem}
where the~CP violating phase $\varphi$ is defined by
$${\rm Im}\,[m_\tau
V_{e\tilde{\tau}}^{{\rm e}_L}
V_{e\tilde{\tau}}^{{\rm e}_R}
V_{\tau\tilde{\tau}}^{{\rm e}_L*}
V_{\tau\tilde{\tau}}^{{\rm e}_R*}] \equiv
|m_\tau
V_{e\tilde{\tau}}^{{\rm e}_L}
V_{e\tilde{\tau}}^{{\rm e}_R}
V_{\tau\tilde{\tau}}^{{\rm e}_L*}
V_{\tau\tilde{\tau}}^{{\rm e}_R*}|\sin\varphi$$
and
$V^{{\rm e}_R}$ ($V^{{\rm e}_L}$) are
the left (right) handed lepton-slepton mixing angles
which appear at the gaugino vertices in the mass eigenstate basis for
leptons and sleptons.
With our assumptions the lepton-slepton mixing angles are
linked to the CKM ones in the following way
\begin{equation}
|V_{e_i\tilde{\tau}}^{{\rm e}_{L,R}}(M_Z)| =
y_t y_b y_\tau^{p_{L,R}} |V_{td_i}(M_Z)|\times\frac
{m^2_{\tilde{e}_{L,R}}(M_{\rm G})-m^2_{\tilde{\tau}_{L,R}}(M_{\rm G})}
{m^2_{\tilde{e}_{L,R}}(M_Z)-m^2_{\tilde{\tau}_{L,R}}(M_Z)},
\qquad i=1,2.
\end{equation}
where $p_L=1$ and $p_R=2$.
The dominant constraint is today given by the
$\mu\to e\gamma$ decay, for which, at present,
${\rm B.R.}(\mu\to e\gamma)<4.9 \cdot 10^{-11}$~\cite{Botton}.
The experimental study of
$\mu\to e$ conversion, currently limited by
${\rm C.R.}(\mu\to e~{\rm in~Ti})< 10^{-12}$~\cite{SINDRUM},
 may undergo a very significant
progress in the near future~\cite{MuEConvExp}.
For large values of
the CP violating phase $\varphi$, the bound on the
electric dipole moment of the electron,
$|d_e|<4.3\cdot 10^{-27}\,e\cdot{\rm cm}$~\cite{CRDR} gives the
same restriction in parameter space as $\mu\to e\gamma$.

The dominant amplitude $F$, given by Feynman graphs containg
both left-handed and right-handed sleptons\footnote{With
`left-handed slepton' we mean the supersymmetric partner of the
corresponding left-handed lepton.}, is proportional to the left-right
slepton mixing term in the Lagrangian, which gets contributions
proportional either to the $A$-terms or to $\mu\tan\beta$.
In view of $\tan\beta=45\div60$,
we assume that
$\mu\tan\beta\gg A$ and we consequently neglect the $A$-terms
contributions\footnote{The variations of the $A$ terms in their
allowed range has actually little influence on the determination
of allowed region of the $\{M_2,\mu\}$ plane and on
the $\mu \to e \gamma$ amplitude.}.
We are also neglecting LFV mediating diagrams which employ only
right-handed (or left-handed) sleptons because of a
$(m_\mu/m_\tau)$ suppression factor~\cite{8Art}. One may wonder
whether this is plausible since, due
to the ${\rm U}(1)_X$ $D$-term, the $\tilde{\tau}_L$ may be
significantly lighter than the $\tilde{\tau}_R$. But these
graphs are indeed negligible because the graphs with left-right
slepton mixing have an
additional enhancement factor $\tan\beta$.
It is therefore possible to approximate the amplitude with\footnote{The
contribution from the $m_\tau\mu\tan\beta$
$LR$-mixing term to the $\tilde{\tau}$ masses is negligible if the
sleptons are heavy enough so that the experimental upper bounds
on the LFV effects are not exceeded.}
\begin{equation}
F =\mu\tan\beta
[G_2(m_{\tilde{\tau}_L}^2,m_{\tilde{\tau}_R}^2)-
G_2(m_{\tilde{e}_L}^2,m_{\tilde{\tau}_R}^2)-
G_2(m_{\tilde{\tau}_L}^2,m_{\tilde{e}_R}^2)+
G_2(m_{\tilde{e}_L}^2,m_{\tilde{e}_R}^2)],
\end{equation}
where
\begin{eqnarray*}
G_2(m^2) &=& \sum_{n=1}^4 \frac{H_{n\tilde{B}}}{M_{N_n}}
(H_{n\tilde{B}}+\cot\theta_{\rm W}H_{n\tilde{W}_3})
\cdot g_2(\frac{m^2}{M_{N_n}^2}),\\
G_2(m_1^2,m_2^2) &=& \frac{G_2(m_1^2)-G_2(m_2^2)}{m_1^2-m_2^2},\qquad
g_2(r)={1\over2(r-1)^3}[r^2-1-2r\ln r].
\end{eqnarray*}
In this equation
$N_n$, $n=1,\ldots,4$ are the four neutralino mass eigenstates,
of mass $M_{N_n}$, related to the bino and the neutral wino by
\begin{equation}
\begin{array}{c}
\tilde{B}  = \sum_{n=1}^4 N_n H_{n\tilde{B}}, \\[2mm]
\tilde{W}_3= \sum_{n=1}^4 N_n H_{n\tilde{W}_3}.
\end{array}
\end{equation}
So, the only low energy parameters on which $F$,
and consequently the LFV processes, depend are
\begin{itemize}
\item the lepton-slepton mixing angles;
\item the $\mu$ parameter and the neutralino masses. Using the GUT
relation $M_1\simeq M_2\cdot \alpha_1(M_Z)/\alpha_2(M_Z)$, all the
neutralino masses may be
computed in terms of $M_2$ and $\mu$, that we take as free parameters;
the sign of $\mu$ turns out to be an irrelevant parameter in almost all
of the parameter space.
\item the slepton masses: for a given value of the right-handed
selectron mass the other slepton masses may be computed as function of
$m_{\tilde{e}_R}$, $M_2$, $\mu$, and of the top Yukawa coupling via the
Planck-scale universality hypothesis. We forget the weak dependence on
the $A$-terms.
\end{itemize}
We present now the contour-plots of the
${\rm B.R.}(\mu\to e\gamma)$ in the plane $\{M_2,\mu\}$ for
$|V_{ts}(M_Z)| = 0.04$, $|V_{td}(M_Z)| = 0.01$ and
at fixed values of the right-handed selectron mass
$m_{\tilde{e}_R}=1\TeV$, and of the top quark Yukawa coupling at the
unification scale, $\lambda_{\rm G}$.
We have explained in the previous section
why we choose to restrict the plane to low $M_2$ and $\mu$ values,
and why the allowed range is a strip.
The $A$-terms and the sign of $\mu$ are additional but irrelevant
parameters.

In fig.~\ref{fig:LFV} we show our predictions for
${\rm B.R.}(\mu\to e\gamma)$
in the allowed area of the $\{M_2,\mu\}$ plane,
for $m_{\tilde{e}_R}=1\TeV$ and
$\lambda_{\rm G} = \{0.85,1\}$.
The allowed area is limited from below
by the $m_A^2>0$ condition (dashed line)
and from above by $m_{\tilde{b}_R}^2>0$ (dashed line)
and $m_{\tilde{L}_3}^2>0$ (dot-dashed line).
The bound at small $M_2$ (dotted line) is obtained by
requiring the charginos to be heavier than $45\GeV$.
The plane is also restricted to values of $M_2$ and $\mu$
lower than $m_{\tilde{e}_R}$
for which the necessary fine-tuning is less severe.

Let us now comment on the dependence of the LFV rates
on the top quark Yukawa coupling.
Its role consists in giving rise to a splitting between the
$\tilde{\tau}$ and $\tilde{e}$, $\tilde{\mu}$ masses
proportional to $\lambda_{t\rm G}^2$.
This splitting is
essential to undo a GIM-like cancellation in the relevant amplitudes.
In SO(10) unified models
we expect that the LFV amplitudes increase as $\lambda_{t\rm G}^4$,
since the dominant LFV-mediating Feynman graphs
are zero unless the non-degeneracy is present
in both the left {\em and} in the right slepton sector.
However, as discussed in the previous section,
in the large $\tan\beta$ region
the parameter space closes around
$\lambda_{\rm G}\approx 1.2$,
when the Higgs pseudoscalar,
the right-handed sbottom and
the left-handed stau
become too light.
For this reason the left-handed stau is
generically lighter for larger values of $\lambda_{\rm G}$.
This give rise to a dependence
of the LFV rates on $\lambda_{\rm G}$
stronger than the naive expectation.
This can be seen in fig.~(\ref{fig:BRlG}),
where the ${\rm B.R.}(\mu\to e\gamma)$
is plotted as function of $\lambda_{\rm G}$
for $m_{\tilde{e}_R}=1\TeV$ and
for various spectra of the supersymmetric particles with $\mu<0.5\TeV$.
The loop integral $F$
turns out to be roughly proportional to $\lambda_{\rm G}^8$.

The rates for the other interesting LFV processes
may also be easily deduced from the graphs, since they are connected to
$\mu\to e\gamma$ in the following way:
\begin{equation}\label{eq:BRCaptDec10N}
\frac{{\rm B.R.}(\mu\to e\gamma)}{5\cdot 10^{-11}}\approx 4
\frac{{\rm C.R.}(\mu\to e{\rm~in~Ti})}{10^{-12}}\sim \left(
{|d_e|/\sin\varphi\over 10^{-26}\,e\cdot{\rm cm}}
\right)^{\!\!2}
\end{equation}
For other values of the right-handed selectron mass than the one
considered, the factor $F$ scales as $(1\TeV/m_{\tilde{e}_R})^{2}$.

The main result is that, in the large $\tan\beta$ region, the present
experimental limits on LFV already constrain in a significant way the
parameter space. More importantly, the experiments in progress and/or
foreseen will be able to probe the theory up to $\tilde{e}$ and
$\tilde{\mu}$ masses of about $1\TeV$ in all of the significant region
of the remaining parameter space.

\mysection{More general and realistic models}\label{TeXtures}
In this section we discuss the problem of how we
may expect the mixing angles in the lepton Yukawa coupling matrix
to be linked to the measured CKM angles by unification physics.

We assume, as explained before, that at the
unification scale all the significant flavour violating terms
are contained in the three Yukawa coupling matrices
$\mb{\lambda}^{\rm a}$, where ${\rm a} = \{{\rm u,d,e}\}$, defined by
$$f_{\rm MSSM} =
h^{\rm u} u_R \mb{\lambda}^{\rm u} Q +
h^{\rm d} d_R \mb{\lambda}^{\rm d} Q +
h^{\rm d} L   \mb{\lambda}^{\rm e} e_R + \mu h^{\rm u}h^{\rm d}.$$
In order to find the physical flavour-violating angles, we
parametrize the Yukawa coupling matrices in the usual way
$$\mb{\lambda}^{\rm a} = \mb{U}^{\rm a\dagger}\cdot
\diag(\lambda_{\rm a_1},\lambda_{\rm a_2},\lambda_{\rm a_3})
\cdot \mb{V}^{\rm a}$$
and rotate the fermion fields to their mass eigenbasis.
Other than the quark-squark mixing angles and the CKM angles,
given by
$\mb{V} = \mb{V}^{\rm u}\cdot \mb{V}^{\rm d\dagger}$,
we are left with the lepton-slepton mixing matrices
at the gaugino vertices
$${\cal L} \supset
(\bar{\lambda}\bar{e}_R \mb{V}^{{\rm e}_R} \tilde{e}_R +
\tilde{e}_R^* \mb{V}^{{\rm e}_R\dagger} e_R\lambda) +
(\lambda L \mb{V}^{{\rm e}_L} \tilde{L}^* +
\tilde{L} \mb{V}^{{\rm e}_L\dagger} \bar{L}\bar{\lambda})$$
with, in the left sector $\mb{V}^{{\rm e}_L} = \mb{U}^{\rm e}$
and, in the right sector $\mb{V}^{{\rm e}_R} = \mb{V}^{\rm e}$.
The symbol $\lambda$ indicates the $\SU(2)_L\otimes{\rm U}(1)_Y$
gauginos.

In models where the up-quark Yukawa couplings are not unified with those
of the down-quarks and of the leptons (because either the fermions or
the two Higgs doublets are not contained in a single
representation of the unification gauge group) the only problem is to
find the link between $\mb{\lambda}^{\rm e}$ and $\mb{\lambda}^{\rm d}$.
In fact, due to the sufficiently large flavour symmetry, the
up-quark Yukawa matrix may be diagonalized at the Planck scale,
so that the CKM mixing angles are entirely contained in
$\mb{\lambda}^{\rm d}$.
In this case the only problem is to connect the
lepton mixing angles with the down-quark ones.

In our SO(10) model, where also the up-quark Yukawa matrix is unified
with the down-quark and lepton ones, the up-quark mixing angles
can not be freely rotated away and
give rise to $u\tilde{u}$ flavour mixing matrices at the gaugino vertices.
So, the full unification of the Yukawa couplings
poses the further problem of separating the contributions to
CKM mixing angles from the up and the down quark mixing matrices.
However the full unification also allows for more predictions
in the Yukawa sector.
This issue has already been studied.
In particular, some work has been done in trying to discover
phenomenologically acceptable textures
of the Yukawa coupling matrices
which can be justified by the physics at the unification scale.
The aim of these works~\cite{GY,ARDHS} consists in studying whether
unified theories may be predictive in the flavour sector.
For this reason, textures with a minimal number of non zero entries
were constructed, which can accommodate the observed spectrum of
lepton and quark masses and CKM mixing angles
in terms of a reduced set of free parameters.
Since we are assuming that the sfermion mass matrices
at the unification scale
are diagonal but not flavour degenerate,
these textures give us also a prediction of how the CKM angles are
connected with the angles in the leptonic sector
which give rise to the LFV processes.

Of some interest are textures of the form~\cite{ARDHS}
$$
\mb{\lambda}^{\rm u} = \lambda_{\rm G}
\pmatrix{0&C_u&0\cr C_u &0E&c_{\rm u}^L B \cr 0& c_{\rm u}^R B &1},
\qquad\mb{\lambda}^{\rm d} = \lambda_{\rm G}
\pmatrix{0&1C&0\cr1C&1E&c_{\rm d}^L B\cr 0 & c_{\rm d}^R B &1},
\qquad\mb{\lambda}^{\rm e} = \lambda_{\rm G}
\pmatrix{0&1C&0\cr1C&3E&c_{\rm e}^R B\cr 0 & c_{\rm e}^L B & 1}$$
where $\lambda_{\rm G}$, $B$, $C$
and $E$ are free parameters
($E$ may be chosen as the only complex one;
$C_u$, not relevant for our considerations,
may be given by a different operator than $C$,
or it may be linked to it by $C_u=-{1\over 27}C$~\cite{ARDHS}),
all the predicted Clebsh factors has been
explicitly written down,
and, in order to obtain acceptable predictions
for $m_c/m_t$ and $V_{cb}$,
only nine distinct possibilities exist for
the Clebsh coefficients $c_{\rm a}^L$ and $c_{\rm a}^R$
of the `23' operator.

The Cabibbo angle is mainly given by $\mb{\lambda}^{\rm d}$
rather than by $\mb{\lambda}^{\rm u}$, and the corresponding angle
in the leptonic sector is $\Frac{1}{3}$ of it.
So, defining
$\chi_{L,R} \equiv c_{\rm e}^{L,R}/(c_{\rm d}^L-c_{\rm u}^L)$,
the Clebsh factors which multiply the rates for the LFV processes
are
\begin{equation}
(\frac{1}{3}\chi_L \chi_R)^2\times\left\{
\begin{array}{l}
{\rm B.R.}(\mu\to e\gamma)\\
{\rm C.R.}(\mu\to e)\end{array}\right.,\qquad{\rm and}\qquad
(\frac{1}{9}\chi_L\chi_R)\times d_e
\end{equation}
In the nine possible cases for the `23' operator
the values of the Clebsh correction factors for
the $\mu\to e\gamma$ and $\mu\to e$ rates
range around one in the interval $10^{-1}\div 10$
with the exception of two models
in which they are around $10^{\pm 2}$.
The constant suppression factor, $1/9$, due to the
Georgi-Jarlskog factor~\cite{GY}
is generally compensated by the other factors.
For the electric dipole moment of the electron
the Clebsh correction factors range from $10^{-1}\div 1$.

At least based on the examples considered, we conclude that
neglecting all Clebsh factors (assumption 8.\ of this paper)
gives, with an uncertainty of one order of magnitude,
a correct estimate of their effects in the LFV rates,
and probably a  slight over-estimate of the electric dipole moment
of the electron by a factor 3.

\mysection{Conclusion}
In supersymmetric theories with no conservation of
lepton flavour, processes like $\mu\to e\gamma$,
$\mu\to e $ conversion or similar may play a very important role.
We have computed the rates for lepton
flavour violating processes in a
supersymmetric~SO(10) model with large
$\tan\beta$ and Yukawa coupling unification
$\lambda_t=\lambda_b=\lambda_\tau$.
These rates are the minimal ones we expect to be present
in more general gauge and Yukawa unified models.
The main result is that, in the large $\tan\beta$ region, the present
experimental limits on LFV already constrain in a significant way the
parameter space. More importantly, the LFV experiments in progress
and/or foreseen will be able to probe the theory up to $\tilde{e}$ and
$\tilde{\mu}$ masses of about $1\TeV$ in all of the significant region
of the remaining parameter space.

\paragraph{Acknowledgements} ~

\noindent The authors are grateful to Riccardo Barbieri
for many discussions and for a careful reading of the manuscript.

\appendix
\mysection{Renormalization from $M_{\rm Pl}$ to $M_{\rm G}$}
Neglecting all couplings except the gauge and the third generation
unified Yukawa one, the solutions to all the one loop RGEs between
$E_{\rm max}=M_{\rm Pl}$ and
$E_{\rm min}=M_{\rm G}$ can be given analytically.

All the equations and their solutions may easily be adapted
from the ones of the model defined in eq.~(22) and
discussed in appendix~A of~\cite{8Art}, provided
that the 10-plet `$\Phi_{\rm d}$', its couplings and its $A$-terms
are erased and
that the coupling named `$\lambda_t$' is identified with
the unified third generation coupling $\lambda$.
The transcription is made easier by the fact that the same notations
and the same values of the high energy parameters
have been used in this article.

\begin{table}
$$\begin{array}{|c||c|ccccc|ccc|}\hline
i&\phantom{-}b_i&c_i^Q&c_i^u&c_i^d&c_i^L&c_i^e&\vphantom{X^{X^X}}
c_i^{\rm u}&c_i^{\rm d}&c_i^{\rm e}\\[0.5mm] \hline
1&{33\over5}&{1\over30}&{8\over15}&{2\over15}&\vphantom{X^{X^X}}
{3\over10}&{6\over5}&{13\over15}&{7\over15}&{9\over5}\\
2&1&{3\over2}&0&0&{3\over2}&0&3&3&3\vphantom{X^{X^X}}\\
3&-3&{8\over3}&{8\over3}&{8\over3}&0&0&{16\over3}&{16\over3}&0
\vphantom{X^{X^X}}\cr\hline
\end{array}$$
\caption{Values of the RGE coefficients in the MSSM.\label{Tab:bcude}}
\end{table}

\mysection{Renormalization from $M_{\rm G}$ to $M_Z$}
Neglecting all couplings except the gauge
$g_i$ ($i=\{1,2,3\}$) and the third generation Yukawa ones
$\lambda_{\rm a_3} = \lambda_a$ (${\rm a} = \{{\rm u,d,e}\}$
and $a \equiv {\rm a}_3=\{t,b,\tau\}$),
the one loop RGEs between
$M_{\rm G}$ and $M_Z$ are~\cite{Rattazzi,RGE}
\begin{eqnsystem}{sys:eqs}
\frac{d}{dt} \frac{1}{g_i^2} &=& b_i,\qquad
\frac{d}{dt} \frac{M_i}{g_i^2} = 0\\
\frac{d}{dt}\lambda_{{\rm a}_g}^2 &=& \lambda_{{\rm a}_g}^2
(c_i^{\rm a} g_i^2 - S_{{\rm a}_g b}^{~} \lambda_{b}^2)\\
\frac{d}{dt}A_{{\rm a}_g} &=& c_i^{\rm a} g_i^2 M_i -
S_{{\rm a}_g b}^{~} \lambda_{b}^2 A_{b}\\
\frac{d}{dt}\mu &=&{\textstyle\frac{1}{2}}
(2c_i^h g_i^2 - S_b \lambda_b^2)\mu\\
\frac{d}{dt} B &=& 2c_i^h g_i^2 M_i -S_b \lambda_b^2 A_b\\
\frac{d}{dt} m_{R_g}^2 &=& 2c_i^R g_i^2 M_i^2  -
\delta_{g3} Z_{Rb}  \lambda_{b}^2 (X_{b} + A_{b}^2) -
{\textstyle\frac{3}{5}} Y_R g_1^2  X_Y
\end{eqnsystem}
where the index $g = \{1,2,3\}$ runs over the generation number,
$R$ runs over the scalar fields of the MSSM and
$Y_R$ is the hypercharge of the representation $R$.
The values of the numerical coefficients $b_i$ and $c_i$
are given in table~\ref{Tab:bcude} while
$Z_{Ra}$ is the coefficient of
the contribution from the Yukawa coupling
$\lambda_a^2$ to the wave-function renormalization of the field $R$
 --- it is different from zero only for the third generation particles
and the for the two Higgs doublets.
Its values are
$$Z^T =
\bordermatrix{&h^{\rm d} & h^{\rm u}
&\tilde{Q}_3&\tilde{t}_R&\tilde{b}_R&\tilde{\tau}_R&\tilde{L}_3\cr
t&0&3&1&2&0&0&0\cr b&3&0&1&0&2&0&0\cr\tau&1&0&0&0&0&2&1\cr}.$$
Due to the non renormalization-theorem,
the running of the Yukawa couplings and of the $\mu$ term
is only induced by
wave function renormalizations.
Then $S_b=\sum_{R=h^{\rm u},h^{\rm u}} Z_{Rb} = (3,3,1)$ and
$S_{{\rm a}_g b} = \sum_{R({\rm a}_g)} Z_{Rb}$
where the sum is extended over the fields involved in the Yukawa
coupling $\lambda_{{\rm a}_g}$, that is
$$S_{ab} = \pmatrix{6&1&0\cr 1&6&1\cr 0&3&4\cr}\qquad\hbox{and}\qquad
S_{{\rm a}_1 b} =S_{{\rm a}_2 b} =
\pmatrix{3&0&0\cr 0&3&1\cr 0&3&1\cr}.$$
The $X$'s are linear combination of sparticle masses defined by
$X_a = \sum_{R(a)} m_R^2$ and $X_Y={\rm Tr}\, Y_R^{~} m_R^2$, or,
more explicitely, by
$$X_t =m_{h^{\rm u}}^2 + m_{\tilde{Q}_3}^2 + m_{\tilde{t}_R}^2,\qquad
  X_b =m_{h^{\rm d}}^2 + m_{\tilde{Q}_3}^2 + m_{\tilde{b}_R}^2,\qquad
X_\tau=m_{h^{\rm d}}^2 + m_{\tilde{L}_3}^2 + m_{\tilde{\tau}_R}^2,$$
$$X_Y = (m_{h^{\rm u}}^2-m_{h^{\rm d}}^2)+
\sum_g (m_{\tilde{Q}_g}^2-m_{\tilde{L}_g}^2-
2m_{\tilde{u}_{Rg}}^2+m_{\tilde{d}_{Rg}}^2+m_{\tilde{e}_{Rg}}^2).$$
Parametrizing the solutions for these combinations in terms
of $I_a$ and $I_Y$
$$\begin{array}{lll}
\frac{d}{dt} X_a = 2c_i^{\rm a} g_i^2 M_i^2 -
S_{ab}^{~}\lambda_b^2 (X_b+A_b^2)\qquad &\Rightarrow\qquad&
X_a = X_{a\rm G} + x_2^a M_{5\rm G}^2 - 3S_{ab}I_b\\
\frac{d}{dt} X_Y= b_1 g_1^2 X_Y &\Rightarrow&
X_Y = X_{Y\rm G}(1-f_1^{-1}) \equiv X_{Y\rm G} - I_Y
\end{array}$$
we may then write the following formal solutions for all the equations
\begin{eqnsystem}{sys:sols}
g_i(M_Z) &=& g_{5\rm G}/f_i^{1/2}\qquad
M_i(M_Z) = M_{5{\rm G}}\cdot f_i^{-1}(E)\\
\lambda_{{\rm a}_g}(M_Z) &=& \lambda_{{\rm a}_g}(M_{\rm G})
E_{\rm a}^{1/2} \cdot{\textstyle \prod_b} y_b^{S_{{\rm a}_g b}}\\
A_{{\rm a}_g}(M_Z) &=& A_{{\rm a}_g}(M_{\rm G}) +
x_1^{\rm a} M_{5\rm G}^2 - S_{{\rm a}_g b} I'_b\\
\mu(M_Z) &=& \mu(M_{\rm G})\cdot
E_h\cdot{\textstyle\prod_b} y_b^{S_b}  \\
B(M_Z) &=& B(M_{\rm G}) + 2 x_1^h M_{5\rm G}  - S_b I'_b\\
m^2_{R_g}(M_Z) &=& m^2_{R_g}(M_{\rm G})+ x_2^R
M_{5\rm G}^2 -\delta_{g3} Z_{Rb} 3I_b +
{\textstyle\frac{3}{5} \frac{Y_R}{b_1}} I_Y
\end{eqnsystem}
where $f_i(E)\equiv1+b_i g_{\rm G}^2 t(E)$,
$$
E_\alpha \equiv\prod_{i=1}^3 f_i^{c_i^\alpha/b_i}\qquad{\rm and}\qquad
x_n^R \equiv \sum_{i=1}^3{c_i^R\over b_i}
[1-f_i^{-n}(M_Z)].$$
The parameters
$$y_a(\lambda_{\rm G})
\equiv\exp\left[-{1\over2}\int \lambda_a^2\,dt\right],$$
$I_a$ and $I'_a$ contain {\em all} the
the Yukawa couplings induced corrections.

The CKM matrix elements between a
light and the heavy generation evolve as
$V_{i3}(M_Z)=V_{i3}(M_{\rm G})/y_t y_b$, while
the elements between the light generations
are not renormalized.

The problem of solving the full set of equations is now reduced
to the one of finding solutions for the
parameters $y_a$, $I_a$ and $I'_a$.
This can be done both numerically and analytically.

\begin{figure}[t]
\setlength{\unitlength}{1cm}
\begin{center}
\begin{picture}(18,10.5)
\put(-1.8,-1.4){\includegraphics{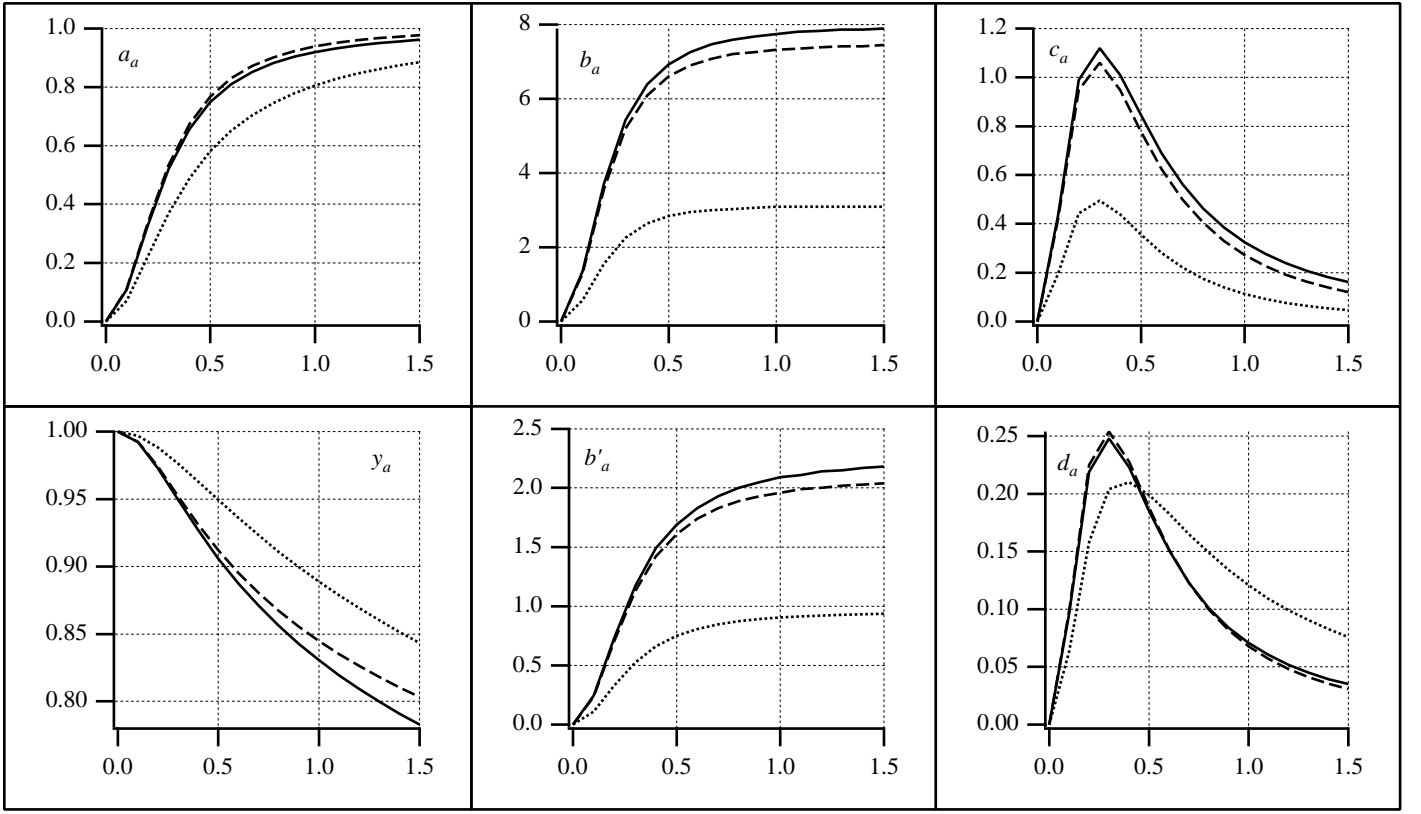}}
%\put(-1,0){\special{picture abcd}}
\end{picture}
\end{center}
\caption{Values of $a_a$, $b_a$, $c_a$,
$d_a$, $b'_a$ and $y_a$ as
function of $\lambda_{\rm G}$ in the interval $0\div1.5$.
The index $a$ runs over $t$ (solid line),
$b$ (dashed line) and $\tau$ (dotted line).
\label{fig:abcd}}
\end{figure}

\subsection{Numerical exact solution}
In our SO(10) model the boundary
conditions at the unification scale for the
$X$'s are particularly simple:
$X_{Y\rm G}=4 m_X^2$ gets contributions only from
the ${\rm U}(1)_X$ $D$-terms,
while the three $X_{a\rm G}$ are all equal and not affected by them:
$X_{a\rm G} = X_{\rm G} \equiv 2 m_{\Psi_3}^2 + m_{\Phi}^2$.
For this reason the form of $I_a$ ed $I_a'$ for the most general SO(10)
symmetric boundary conditions may be given in terms of
few function of the unified Yukawa coupling
\begin{eqnsystem}{sys:IeII}
3S_{ab}I_b &=& a_a(\lambda_{\rm G}) X_{\rm G} +
b_a(\lambda_{\rm G}) M_{5\rm G}^2 + c_a(\lambda_{\rm G}) A_{3\rm G}
M_{\rm G} + d_a(\lambda_{\rm G}) A_{3\rm G}^2\\
S_{ab}I'_b &=& a'_a(\lambda_{\rm G}) A_{3\rm G}+
b'_a(\lambda_{\rm G}) M_{5\rm G}
\end{eqnsystem}
It is easy to see that $a_a = a'_a$.
For a given value of $\lambda_{\rm G}$
a very efficient numerical procedure for obtaining the values of
these coefficients consists in
solving numerically the RGEs four
times for four particular choices of $X_{\rm G}$, $M_{\rm G}$ and
$A_{\rm G}$.
Another possibility consists in obtaining the
RGEs for these coefficients and solving them.

In conclusion we have written the most general solution for all the RGE
in terms of 18 numerical functions.
We show them once for all in fig.~\ref{fig:abcd}.
The solution for the most general SO(10)-invariant boundary conditions
are obtainable by inserting their values
in equations~(\ref{sys:sols}) and~(\ref{sys:IeII}).

\subsection{Analytical approximate solution}\label{Ianal}
Analytical approximate solutions are obtained
setting $S_{ab}\approx 7\cdot\diag(1,1,1)$.
Then
$$y_t \approx y_b \approx[1+12\lambda_{\rm G}^2]^{-1/14},\qquad
y_{\tau}\approx[1+4.4\lambda_{\rm G}^2]^{-1/14}.$$
Defining $\rho_a\equiv 1- y_a^{14}$, the approximate
values of the Yukawa couplings at the $Z$ pole are
$\lambda_t(M_Z) \approx \lambda_b(M_Z) \approx 1.05\cdot\rho_t^{1/2}$
and $\lambda_\tau(M_Z)\approx 0.70\cdot\rho_\tau^{1/2}$.
Approximate expressions for the dimensionful parameters are
\begin{eqnsystem}{sys:Iapprox}
3 I_t\approx 3 I_b&\approx&{\textstyle\frac{1}{7}}
\rho_t[X_{\rm G} + (13-5\rho_t) M_{5\rm G}^2
+(1-\rho_t)A_{\rm G}^2 + 4.5(1-\rho_t) A_{\rm G} M_{5\rm G}]\\
3I_\tau &\approx&{\textstyle\frac{1}{7}}
\rho_\tau[X_{\rm G}+ (8-5\rho_\tau) M_{5\rm G}^2
+(1-\rho_t) {\cal O}(A_{\rm G} M_{5\rm G},A_{\rm G}^2)]\\
I'_t \approx I'_b &\approx &{\textstyle\frac{1}{7}}
\rho_t[A_{\rm G}^2 + 2.2 M_{5\rm G}]\\
I'_\tau &\approx&{\textstyle\frac{1}{7}}
\rho_\tau[A_{\rm G}^2 + 1.4 M_{5\rm G}]
\end{eqnsystem}

\subsection{Evolution of the lepton-slepton mixing angles below the
unification scale}\label{VeEvol}
The lepton-slepton mixing may be described by two
matrices of physical mixing angles, $(\mb{V}^L)_{e_i\tilde{e}_j}$
and $(\mb{V}^R)_{e_i\tilde{e}_j}$, that,
in a supersymmetric basis
(where the lepton-slepton-gaugino interactions are flavour blind)
of sleptons mass eigenstates
measure, respectively, the different orientation in flavour space
between the left (right) slepton mass matrix and the left (right) part
of the lepton Yukawa coupling matrix $\mb{\lambda}^{\rm e}$.
In the calculations we have employed a non supersymmetric
physical basis of mass eigenstates for
both the leptons and the sleptons.
In this case the LFV mixing matrices
appear at the lepton-slepton-gaugino vertex.

To consider their evolution it is better to employ a supersymmetric
basis in which the lepton Yukawa coupling matrix is diagonal.
Then it will remain diagonal throughout renormalization. This is due to
the fact that if there are no LFV at the unification scale, the large
$\lambda_\tau$ Yukawa coupling may render the $\tilde{\tau}$'s lighter
than the other sleptons, but does not generate LFV.
The evolution equations for all the small non-diagonal elements
$\delta m^2$ of the left and the right slepton mass matrices
between the third generation and a light one are
$$\begin{array}{l}
\frac{d}{dt} \delta m^2_L     = -\frac{1}{2}\lambda_\tau^2 \delta m^2_L\\[2mm]
\frac{d}{dt} \delta m^2_{e_R} = -~\lambda_\tau^2 \delta m^2_{e_R}
\end{array}
\qquad\Rightarrow\qquad
\begin{array}{l}
\delta  m^2_L(M_Z)     = \delta  m^2_L(M_{\rm G})\cdot y_\tau\\[2mm]
\delta  m^2_{e_R}(M_Z) = \delta  m^2_{e_R}(M_{\rm G})\cdot y_\tau^2
\end{array}.$$
The mixing angles at the gaugino vertices in the lepton
and slepton physical mass basis,
$$V^{{\rm e}_L}_{e_i\tilde{\tau}}=
\frac{(\delta m^2_L)_{\tilde{e}_i\tilde{\tau}}}
{m^2_{\tilde{e}_L} - m^2_{\tilde{\tau}_L}}\qquad{\rm and}\qquad
V^{{\rm e}_R}_{e_i\tilde{\tau}}=
\frac{(\delta m^2_{e_R})_{\tilde{e}_i\tilde{\tau}}}
{m^2_{\tilde{e}_R} - m^2_{\tilde{\tau}_R}},$$
suffer also a reduction due to the increase in mass difference between
the $\tilde{\tau}$s and the other sleptons. However this is compensated
in physical effects by the reduced mass of the stau leptons.

\end{document}